\documentclass[prl, twocolumn,amsmath,amssymb, superscriptaddress,longbibliography, nofootinbib]{revtex4-1}
\usepackage[pdftex,colorlinks=true]{hyperref}
\hypersetup{citecolor = black}
\usepackage{float}
\usepackage{graphicx}
\usepackage{natbib,hyperref}
\usepackage{comment}
\usepackage{amsmath,epsfig,color,amssymb}
\usepackage{array}
\usepackage{bbold}
\usepackage[dvipsnames]{xcolor}
\usepackage[export]{adjustbox}
\usepackage{comment}
\usepackage{cancel}
\newcommand{\beg}{\begin{equation}}
\newcommand{\en}{\end{equation}}

\newcommand{\bq}{\mathbf q}

\newcommand{\br}{\mathbf r}

\newcommand {\tcg}{\textcolor{blue}}

\newcommand \bel  {\begin{align}}
\newcommand \enl  {\end{align}}

\newcommand{\eps}{\varepsilon}

\newcommand{\dg}{^\dagger}
\newcommand{\ket}[1]{|#1\rangle}
\newcommand{\bra}[1]{\langle#1|}

\definecolor{new}{rgb}{.08,.05,.8}

\begin{document}
\author{Ammar Kirmani} 
\affiliation{Theoretical Division, Los Alamos National Laboratory, Los Alamos, New Mexico 87545, USA}
\email{akirmani@lanl.gov}

\author{Benedikt Fauseweh}
\affiliation{Department of Physics, TU Dortmund University, 44227 Dortmund, Germany}
\email{benedikt.fauseweh@tu-dortmund.de}

\author{Jian-Xin Zhu}
\affiliation{Theoretical Division, Los Alamos National Laboratory, Los Alamos, New Mexico 87545, USA}
\affiliation{Center for Integrated Nanotechnologies, Los Alamos National Laboratory, Los Alamos, New Mexico 87545, USA}
\email{jxzhu@lanl.gov}

\title{Laser-driven Ultrafast Dynamics of a Fractional Quantum Hall System}

\begin{abstract}
Fractional quantum Hall (FQH) systems are strongly interacting electron systems with topological order. These systems are characterized by novel ground states, fractionally charged and neutral excitations. The neutral excitations are dominated by a low-energy collective magnetoroton mode. Here we derive and use a quasi-one-dimensional model to investigate the ultrafast nonequilibrium dynamics of a laser-driven FQH system within a two-Landau-level approximation. As opposed to the traditional and synthetic bilayers, our model accounts for interactions where electrons can scatter from one Landau-level to another. By performing exact time evolution of the system, we create an out-of-equilibrium state following the laser pulse that shows rich physics. Our calculations show the presence of non-trivial excited modes. One of these modes is electromagnetically active and represent density oscillations of \emph{magnetoplasmon} mode. Another mode is identified by evaluating the overlap of the initial state and the out-of-equilibrium state following the laser pulse with a quadrupole operator. This mode is analogous to the chiral-graviton mode for FQH systems recently measured in experiments [Nature {\bf 628}, 78 (2024)]. Our results show that a linearly-polarized pulse field can excite the graviton mode when inter-Landau level scattering occurs.
\end{abstract}
\maketitle

{\it Introduction.---} Fractional quantum Hall (FQH) systems are a prototypical example of strongly correlated physical systems with rich novel phenomena like topological order, quantum geometry and excitations that follow fractional statistics \cite{Haldane2011,Wilzeck1983}. The neutral excitations of FQH systems are dominated by a collective coherent oscillation mode usually called Girvin-MacDonald-Platzman magnetoroton mode \cite{GirvinPRB1986,GirvinPRL1985}. In the long-wavelength limit $k\to 0$, the magnetoroton becomes a quadrupole excitation that carries $L=2$ angular momentum along with the excitation of an underlying quantum geometry metric analogous to the passing of gravitational waves in a medium~\cite{Haldane2011}. Another novel consequence of the presence of quantum geometry and topology in FQH systems is the quantization of Hall 
viscosity~\cite{haldanevisc2009,JainVisc2020, RezayiVisc2011}. FQH phase is characterized by the quantization of transverse component of resistivity at low temperatures. This is measured in 2D electrons with density $\rho$ in the presence of high perpendicular magnetic field $B$ at fractional filling $\nu=\rho h/eB$  ($h$ being Planck's constant and $e$ is the electron charge). Among many odd-denominator fractional filling states observed experimentally~\cite{Tsui1982,Tsui1983} and theoretically explained~\cite{Laughlin1983,HaldaneHierarch}, the first observed FQH state at $\nu=\frac{1}{3}$ is of particular interest. 
It was shown that fractional $\frac{1}{3}$ state can host charged excitations like quasi-particle (quasi-hole) of $e/3$ sometimes called \emph{anyons}~\cite{Wilzeck1983}. It was not until recently that experiments performed on this FQH state gave first direct observation of exchange statistics of fractionally charged particles (anyons) \cite{Nakamura2020}. Furthermore, interest in this state is rekindled due to the observation of two $e/3$ quasi-particles scattering into an electron and ``Andreev reflected'' quasi-hole \cite{scatter2023}. Besides its experimental accessibility, this state has also been readily studied theoretically on different geometries \cite{Yoshioka1983,Rezayi1994,NakaMain,Jolicoeur2012}. Moreover, a long wavelength low-energy excitation mode of $\frac{1}{3}$ FQH state has been excited in inelastic linearly-polarized light scattering experiments \cite{inelastic1/3}. This mode was identified later on as a chiral-graviton mode that can be excited through circularly polarized light~\cite{chiralRezayi}. Another experiment has also excited this mode through two-photon processes  in FQH-1/3 state \cite{NatureGraviton}. Despite such experiments, the theoretical understanding of the dynamics of a multi-Landau-level FQH system has been limited.

\begin{figure}
  \includegraphics[width=\linewidth]{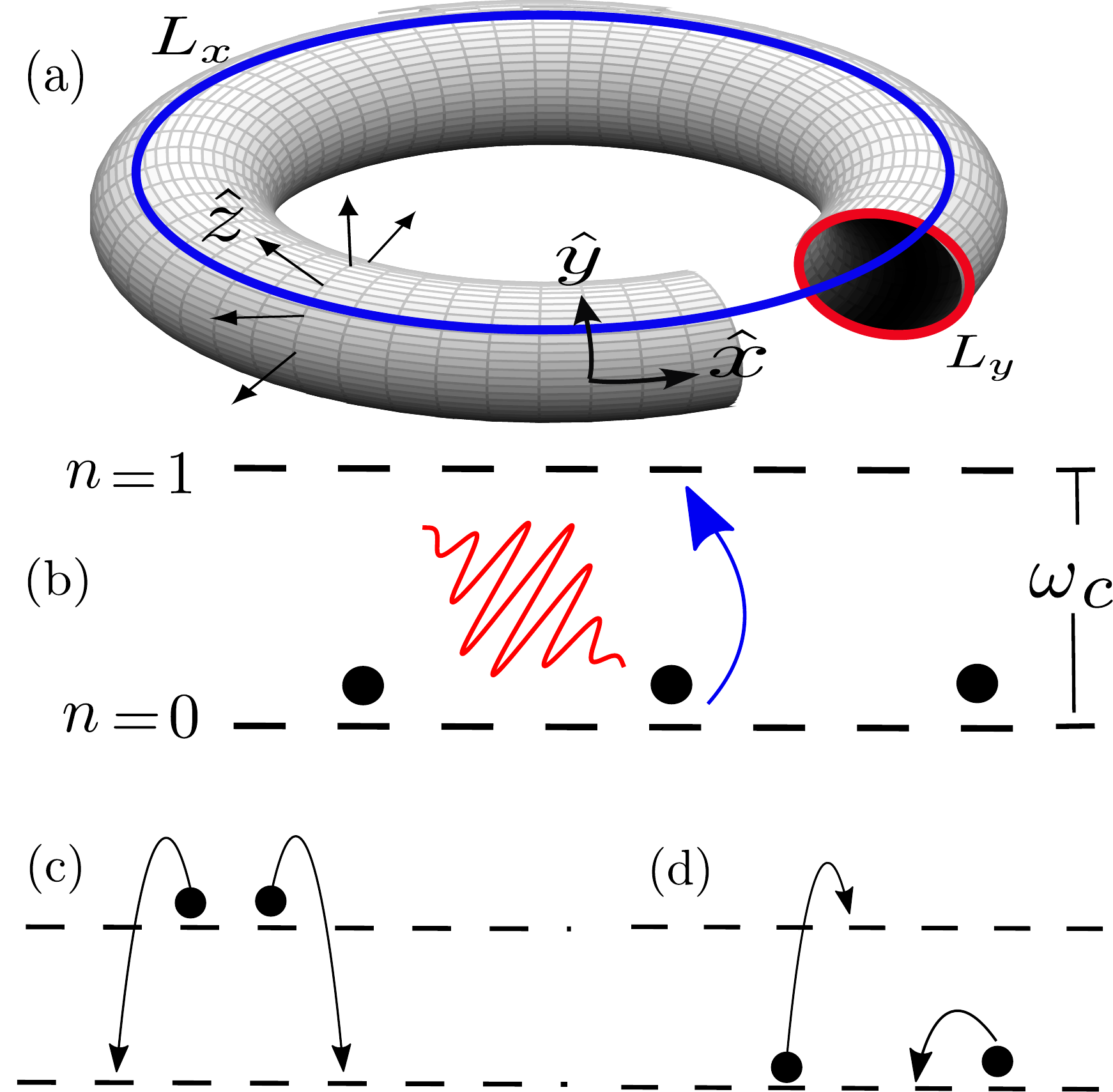}
  \caption{ (a) Torus with lengths $L_x$ and $L_y$. Magnetic field is pointing along $\hat{z}$. (b) A schematic of a two-level FQH system. Initially, the system is in fractional $\nu=1/3$ state occupying the LLL ($n=0$). This state is then laser-driven creating excitations to the first Landau level ($n=1$). The energy difference between the two levels is $\omega_c$. (c) and (d) are the two-electron scattering processes that do not conserve Landau level index. (c) describes two electrons transitioning from one Landau level to another. (d) a two-electron process where only one electron scatters to another Landau level.}
  \label{fig:excite}
\end{figure}

In this Letter, we present ultra-fast dynamics of the FQH system is driven out-of-equilibrium by a laser pulse field. The system is initially in $\nu=\frac{1}{3}$ FQH state. We show that when all inter-two-level scattering terms are considered, a short time linearly-polarized light with frequency near Landau level gap can excite chiral-graviton mode below magnetoplasmon mode. We note that while optically driven synthetic bilayers have been recently proposed~\cite{Pouyan2017,ZeHafezi} to host new types of FQH states~\cite{halp1983,Sarma331}, as in conventional semiconductor bilayers~\cite{biEisen, BiSuen,Bishop,BiAlAs}, the electronic systems are assumed to thermalize to the Floquet ground state in the rotating framework of the optical drive field. Therefore, the problem studied in the present work is quite distinct from the situations studied before.\\
{\it Model Hamiltonian{\color{black}}.---}
We consider a 2D spinless electronic system defined on the surface of torus with lengths $L_x$ and $L_y$, as shown in Fig.~\ref{fig:excite} in the presence of both a high perpendicular static magnetic field and a time-dependent electric field from the laser pulse. In the Landau gauge, both magnetic and electric fields can be described by the vector potential 
$\mathbf{A}=\mathbf{A}_{\rm static}(\mathbf{r}) + \mathbf{A}_{\rm EM}(t)$ with $\mathbf{A}_{\rm static}=(0,Bx,0)$ and $\mathbf{A}_{\rm EM}= (0,A_{\rm EM}(t),0)$. The single-particle states in this gauge compatible with torus boundary conditions are given by~\cite{papicbandmass}
\begin{eqnarray}\label{eq:sp}
\nonumber
\phi_{n,j}(\br)=\frac{1}{\sqrt{{\cal N}_n}}\sum_k &e^{i\frac{(X_j+kL_x)y}{\ell_B^2}}e^{-\frac{(x+X_j+kL_x)^2}{2\ell_B^2}}\\&H_n\bigg(\frac{x+X_j+kL_x}{\ell_B}\bigg)
\end{eqnarray}
Where $X_j=\frac{2\pi j}{L_y}\ell_B^2$ with $\ell_B=\sqrt{\hbar/eB}$ being the magnetic length. ${\cal N}_n=L_y\sqrt{\pi}2^n n!\ell_B$ is the normalization constant and $H_n$ is Hermite polynomial of order $n$. For each Landau level $n$, index $j$ runs from $0,1,...,N_{\phi}-1$, where $N_{\phi}=L_x L_y/(2\pi\ell_B^2)$ is the number of flux quanta (Landau orbitals) passing through the system. Within this Landau orbital basis, we are able to write the system Hamiltonian as $\hat{H} =\hat{H}_{sp} +\hat{H}_{int}$, with 
\begin{eqnarray}\label{eq:Hsp}
 \hat{H}_{sp} &=\omega_c\sum_{j} \hat{n}_{1,j}
+\Gamma_{\rm EM}(t) [\sum_{j} c_{0,j}\dg c_{1,j}+ \mathrm{H.c.}]\;,
\end{eqnarray}
describing the coupling of electrons with the laser pulse field up to the first Landau level (LL1), and 
\begin{equation}
\label{eq:Model}
 \hat{H}_{int} =\sum_{\{n_i,j_i\}}V^{n_1,n_2;n_3,n_4}_{j_1,j_2;j_3,j_4}c_{n_1,j_1}\dg c_{n_2,j_2}\dg c_{n_3,j_3}c_{n_4,j_4}\;,
\end{equation}
describing the Coulomb interaction with $V_{n_1,n_2;n_2,n_4}^{j_1,j_2;j_3,j_4}$ the amplitude of two-particle scattering processes due to the repulsive interactions.
Here the operators $c_{m,j}$, $c_{m,j}\dg$  ($\hat n_{m,j} {\equiv} c_{m,j}^\dagger c_{m,j}$) destroy or create an electron in $m$-th Landau level (LL) at $j$-th orbital localized around $2\pi j \ell_B^2/L_y$ along the axis of the cylinder.  The quantity $\omega_c$ is the level-spacing between Landau levels 0-1 (see Fig. \ref{fig:excite}). In each Landau level, cyclotron frequency of electron is quantized in units of $\omega_c=eB/m^*$ ($m^*$ is electron's effective-mass). The laser field coupling strength $\Gamma_{\rm EM} = \frac{eA_{EM}(t)}{\sqrt{2}m^* \ell_B} $. We also note for electrons interacting through Coulomb repulsion in the system, the Coulomb energy $E_c=e^2/\eps \ell_B$ ($\eps$ being di-electric constant) is also an important energy scale.
 The interaction part of our Hamiltonian Eq.~(\ref{eq:Model})  contains both intra- and inter-level scattering processes.  Throughout the work, we use a short pulse form $A_{\rm EM}(t)=A_0 e^{-\frac{(t-t_c)^2}{2t_d^2}}\cos\big[\omega_0 (t-t_c)\big]$ where $t_d$ is time delay, $t_c$ is the center time for the pulse and $\omega_0$ is the frequency of incoming laser \cite{ZhuKondo}. 

In the absence of the pulse field and for large positive $\omega_c$, our fermions are in the ground state of the system, 
 containing $N$ electrons, such that the filling factor is $\nu{=}N/N_\phi{=}1/3$. Also the system Hamiltonian 
 commutates with the center of mass-momentum operator $\hat{K}=\sum_{j,n} jc_{n,j}\dg c_{n,j}$ (in Landau gauge) and momentum transfer by the laser pulse is considered to be negligible. Hence, our many-body dynamics is contained in the momentum eigen-sector of the initial state. 
To include the effects of the out-of-equilibrium physics under the influence of a laser-pulse, the Hamiltonian includes both scattering processes that conserve Landau level index $n_1+n_2=n_3+n_4$ and ones that break the level index i.e. $n_1+n_2\neq n_3+ n_4$. The latter is schematically illustrated in Fig.~\ref{fig:excite} (c) and (d). This is the consequence of an electron scattering from one level to another and is ignored in traditional and synthetic bi-layers \cite{Pouyan2017}. 
We avoid large scale Hilbert space by truncating Hamiltonian 
to LL1 and by considering low-intensity radiation. This assumption is further validated by another resonant inelastic light experiment for FQH $\frac{1}{3}$ state showing most features related to excitations from LLL to LL1 \cite{inelastic1/3}. Coulumb's interaction potential for torus boundary conditions is $V(\br)=\sum_{s,t}e^2/\eps|\br +s L_x\hat{x}+tL_y\hat{y}|$ where $\hat{x}$, $\hat{y}$ are the units vectors along the two axes of the torus \cite{Yoshioka1983}. We obtain the matrix elements
\begin{widetext}
\begin{eqnarray}\label{eq:H01}
V_{j_1,j_2,j_3,j_4}^{n_1,n_2,n_3,n_4}=\frac{\delta_{j_1+j_2,j_3+j_4}^{'}}{2L_x L_y}\sum_{\bq,\bq\neq 0}\delta_{q_y,t\frac{2\pi}{b}}\delta_{q_x,s\frac{2\pi}{a}}\delta_{j_1-j_4,t}^{'}\frac{2\pi e^2}{\eps q}C_{n_1,n_4}(-\bq)C_{n_2,n_3}(\bq)e^{-i2\pi s\frac{j_1-j_3}{N_\phi}}e^{-\frac{q_x^2+q_y^2}{2}\ell_B^2}
\end{eqnarray}
\end{widetext}
Where $C_{0,0}(\bq)=1$, $C_{1,1}(\bq)=1-\frac{|q|^2\ell_B^2}{2}$ and $C_{1,0}(\bq)=\frac{q_y+iq_x}{\sqrt{2}}\ell_B=\big[C_{0,1}(-\bq)\big]^*$. Prime over a delta function indicates that it is defined modulo $N_\phi$. Fractional $1/3$ state \cite{Laughlin1983} is the ground state of the Hamiltonian $\sum_{j_i}V_{j_1,j_2,j_3,j_4}^{0,0,0,0}c_{j_1}\dg c_{j_2}\dg c_{j_3}c_{j_4}$ and is adiabatically connected to period-three charge-density $\ket{100,100,100,100,...}$ \cite{Seidel2005,Kevilson1985} (for short-range potential $\nabla^2 \delta(\br)$). The physics of FQH systems are usually studied in the limit of large level-spacing  and the Hamiltonian for single Landau level is diagonalized. This is due the assumption that Coulomb interaction is weak to scatter electron from one level to another in the large magnetic field limit which is not always true \cite{papic_semi2022}. We have first verified that the ground state of of the system Hamiltonian 
involving all scattering processes for various values of level spacing is indeed the fractional-$\frac{1}{3}$ quantum Hall state (See Supplemental Material (SM)~\cite{supp} for more details). \\
{\bf \em Dynamical observables.} Our system is initially in FQH-$\frac{1}{3}$ state $\ket{\Psi_0}$ at time $t=0$. We then turn on the laser drive and let the system evolve through Schrodinger equation as $\ket{\Psi(t)}=\hat{\mathit T}e^{-i\int_0^t \hat{H}(t')dt'}\ket{\Psi_0}$, where $\hat{\mathit T}$ is the time ordering operator and $|\Psi(t)\rangle$ is our out-of-equilibrium state. The time-ordering is achieved by discretizing the time-dependent matrix exponential in small time steps. The dynamical response of the system can be understood with time-dependent structure factor ${\mathcal S}(\bq)=\langle\hat{\rho}_{\bq}\hat{\rho}_{-\bq}\rangle$ where $\rho_{\bq}$ is the projected density (see SM ~\cite{supp}). 

{\bf \em Results.}  We consider level-spacing $\omega_c<E_c$, i.e. the level spacing is less than Coulomb energy scale. The ratio of these energies $\kappa\equiv \frac{e^2}{\hbar \omega_c \epsilon\ell_B}$ is magnetic field dependent ranging from $2.6/\sqrt{B[T]}$ for Ga-As to $22.5/\sqrt{B[T]}$ for Al-As justifying the our approximation for fractional $\nu=1/3$ Hall system usually observed at $B \sim 10$ Tesla \cite{mixing2013}.
\begin{figure}
  \includegraphics[width=1.0\linewidth]{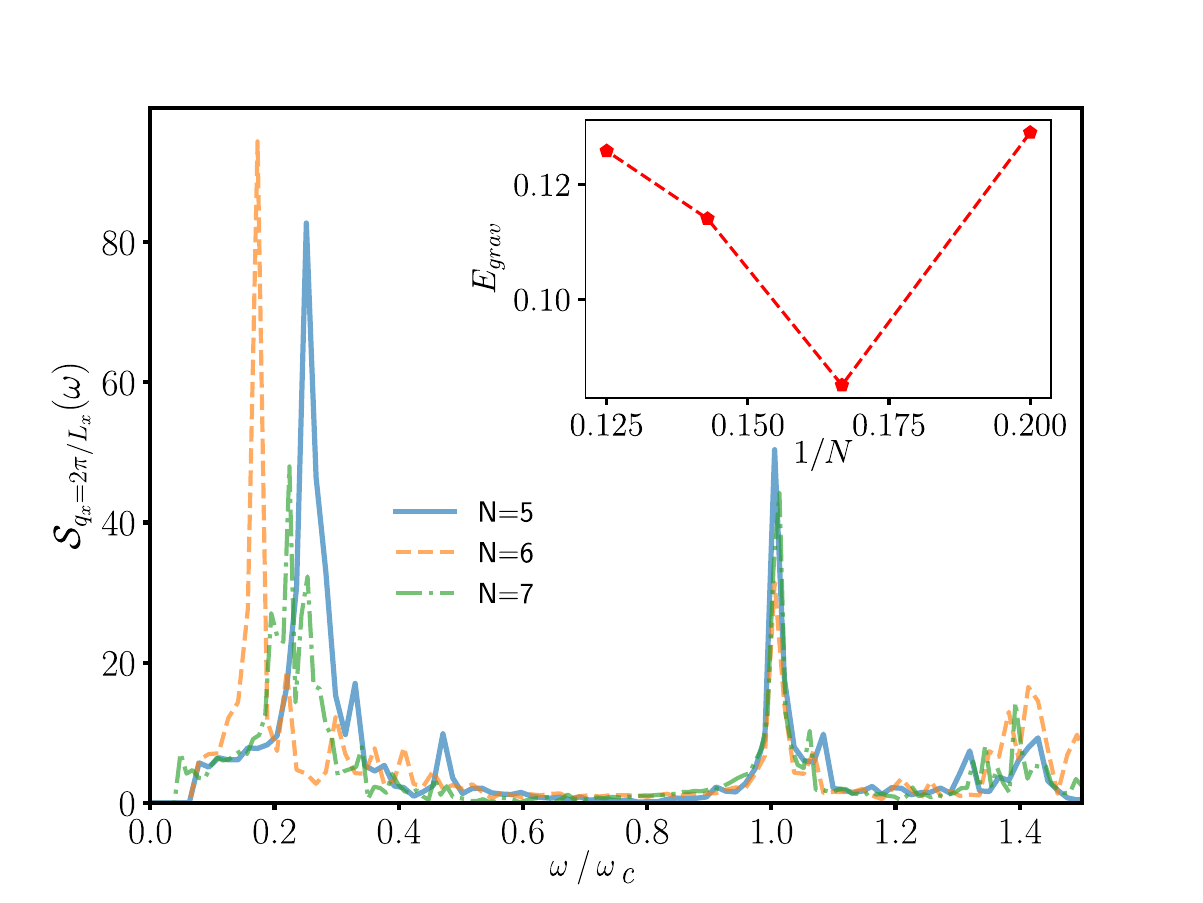}
  \caption{ Color online. Structure factor in frequency domain ${\mathcal S}_{\bq}(\omega)$  for different system sizes. The torus dimensions is taken $L_x=L_y=\sqrt{6\pi N}\ell_B$. The laser-frequency is $\omega_0=0.7\omega_c$. The intensity is set to $eA_0/\sqrt{2}m^*\ell_B=0.2E_c$ and $t_d=10/E_c$. The units of time is the inverse of $E_c$. The usual values of $E_c$ is around $14\; {\rm meV}$ \cite{inelastic1/3} which gives $47$ femto-seconds as the unit of time. The structure factor spectrum is normalized such that $\int {\mathcal S}_{\bq}(\omega)d\omega=1$.}
  \label{fig:sq}
\end{figure}
The spectrum of structure factor ${\mathcal S}_\bq(\omega)$ is shown in  Fig. \ref{fig:sq}. The peak at the level-spacing (cyclotron-energy) $\omega_c$ is the plasmon equivalent of FQH state identified as $q\to0$ limit of \emph{magnetoplasmon} mode (see SM \cite{supp} for other quantities). Significantly, there emerge new modes located at the energy around $0.2 \omega_c$ for different system sizes. To determine the nature of theses new low-energy modes in $S_{\bq}(\omega)$, we evaluate the overlap $|\langle{\Psi(t)}\ket{n}|$ to analyze the weight of low-lying eigen-states $\ket{n}$ contained in the photo-induced excited state $\vert \Psi(t)\rangle$ as shown in Table \ref{tab:grav_updated}. We have found that low-energy excitations are dominated by quadrupole FQH graviton states. These states are identified by matrix-element $I_n=|\bra{\Psi_0} \hat{O}^{(2)}\ket{n}|$, where $\hat{O}^{(2)}$ is the operator with quadrupole structure and $\ket{\Psi_0}$ is our initial ground state before the application of laser pulse. The specific form of $\hat{O}^{(2)}$ is given as~\cite{chiralRezayi}
\begin{eqnarray}\label{eq:grav}
\hat{O}^{(2)}=\sum_{\{j_i\}}\delta_{j_1+j_2,j_3+j_4}^{'}\hat{O}^{j_1,j_2,j_3,j_4}_{qd} c_{0,j_1}\dg c_{0,j_2}\dg c_{0,j_3}c_{0,j_4}\quad\\
\hat{O}^{\{j_i\}}_{qd}=\sum_{q=(s \frac{2\pi}{L_x},t\frac{2\pi}{L_y})}^{'} \delta_{j_1-j_4,t}^{'}\bigg(q_x^2-q_y^2\bigg)\frac{2\pi}{q}e^{-\frac{q^2}{2}}e^{-i2\pi s\frac{j_1-j_3}{N_\phi}}\nonumber
\end{eqnarray}

\begin{figure}
\centering
  \includegraphics[width=1.0\linewidth]{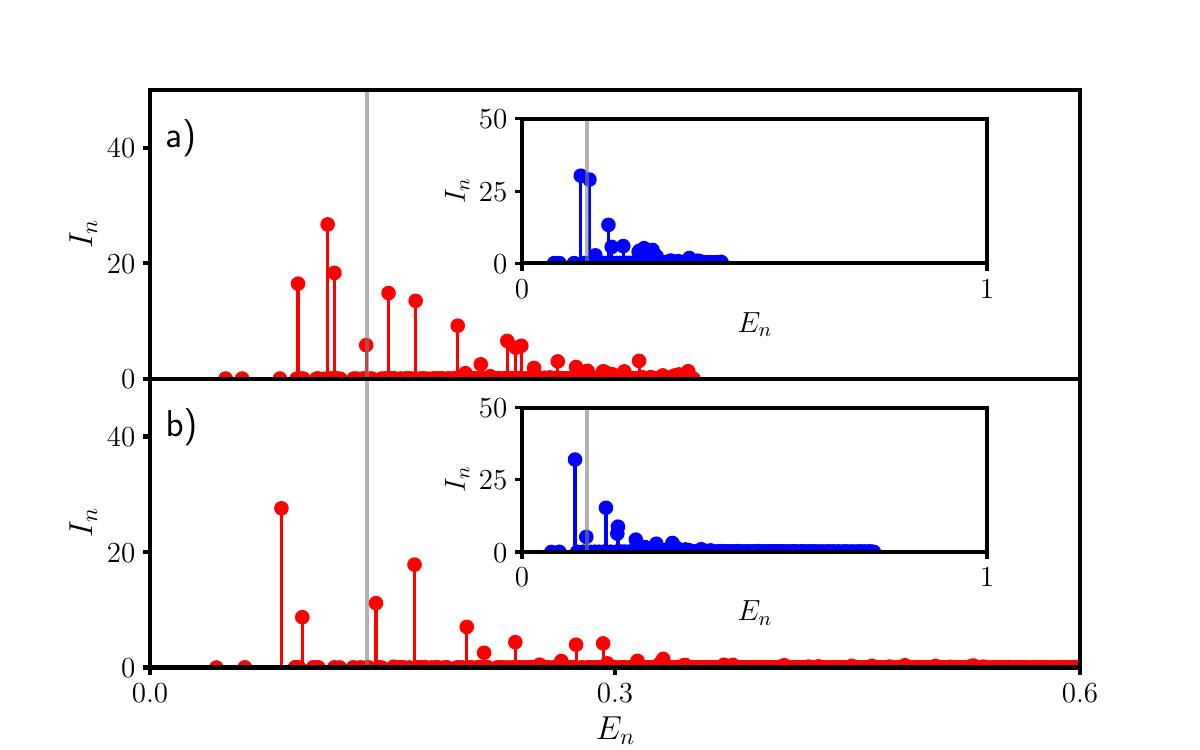}
  \caption{Color online. (a) $I_n=|\bra{\Psi_0} \hat{O}^{(2)}\ket{n}|$ for $N=7$ for $L_y=\sqrt{2\pi 21} \ell_B$. Where $\ket{\Psi_0}$ is the FQH-1/3 state and $\ket{n}$ is the $n$-th excited state of Hamiltonian in Eq. (\ref{eq:Model}). The inset of (a) gives the transition matrix element between FQH-1/3 state and the eigen-states of LLL. (b) Same as (a) but for $N=6$ and $L_y=\sqrt{2\pi 18}\ell_B$. The level-spacing is set to $0.5E_c$. Grey vertical line on plots is at energy $0.14 E_c$. }
  \label{fig:grav}
\end{figure}
\begin{table}\label{tab:grav_updated}
\caption{\label{tab:grav_updated} The contribution of states in the photo-induced state $\ket{\Psi(t)}$ after the pulse for $N=$5, 6, 7 and 8. $\Delta E$ denotes the energy difference from the ground state. Overlap $|\langle n\ket{\Psi(t)}|$ gives the contribution of individual eigenstate $\ket{n}$ of the system Hamiltonian with energy $E_n$. $\omega_c=0.5E_c$.}
\begin{ruledtabular}
\begin{tabular}{ccccc}
 Size \footnotemark[1] &$E_n$ ($E_c$) & Overlap &$\Delta E$ $(E_c)$ & Nature  \\
 \hline
 N=5 & -1.067& 0.883 & 0.000 & G.S.\\
 -& -0.938& 0.097& 0.129 & Quadrupole\\
  -& -0.852& 0.01& 0.215 & Quadrupole\\

 -& -0.562& 0.4 & $0.504\approx \omega_c $ & Plasmon\\
 \hline
 Total weight&& 0.097&&Quadrupole   \\
\hline
$N=6$ &-1.381 & 0.850 & 0.000 &G.S.\\
-& -1.296 & 0.095 & 0.085& Quadrupole\\
-&-1.283& 0.041 & 0.098 & Quadrupole \\
-&-1.235& 0.028 & 0.146 & Quadrupole \\
-&-1.210&0.034& 0.171& Quadrupole\\
-&-1.176&0.014& 0.205 & Quadrupole\\
-&-0.876& 0.429 & $w_c$ & Plasmon \\
\hline
Total weight&&0.212&&Quadrupole\\
\hline
 N=7 \footnotemark[2] & -1.702& 0.827 & 0.000 & G.S.\\
 -& -1.606& 0.058& 0.096 & Quadrupole\\
 -& -1.587& 0.065 & 0.115  & Quadrupole\\
 -& -1.583& 0.046 &  0.119 & Quadrupole\\
  -& -1.562& 0.013 &  0.140 & Quadrupole\\
 -& -1.548& 0.030 &  0.161 & Quadrupole\\
  -& -1.530& 0.024 &  0.172& Quadrupole\\
 -& -1.2 & -& $w_c$& Plasmon\\
 \hline
Total weight&&0.236&& Quadrupole \\
\hline
 N=8 \footnotemark[2] & -2.035& 0.797 & 0.000 & G.S.\\
 -& -1.951 & 0.026& 0.084 & Quadrupole\\
 -& -1.926& 0.049 & 0.109  & Quadrupole\\
 -& -1.909&  0.063 &  0.126 & Quadrupole\\
  -& -1.896&  0.013 &  0.139 & Quadrupole\\
 -& -1.884& 0.038 &   0.151 & Quadrupole\\
  -& -1.873& 0.019 &   0.162& Quadrupole\\
 \hline
Total weight&&0.208&& Quadrupole
\end{tabular}
\\
\end{ruledtabular}
\footnotetext[1]{The parameters of the laser pulse are set to $(eA_0/\sqrt{2}m^*\ell_B,\omega_0)=(0.2E_c,0.7\omega_c)$ with $t_d=10E_c$ and $L_y=\sqrt{2\pi N_\phi}\ell_B$. Similar results have been found for observed for near-TT limit.}
\footnotetext[2]{Due to large Hilbert space, for $N=7$, we only have low-energy quadrupole states by which we identify graviton modes in the structure factor. For $N=8$, we have only have eigenvectors corresponding to first 100 lowest energy eigenstates that include few graviton states. It is expected that for $N=8$ 
States that give more than $0.01$ overlap with $|\psi(t)\rangle$ are considered.} 
\end{table}
In Fig. \ref{fig:grav}, we present the values of transition matrix element $\bra{n}\hat{O}^{(2)}\ket{\Psi_{0}}$ (un-normalized) between FQH 1/3 state and  $n$-th excited state $\ket{n}$ for two cases i) When $\ket{n}$ is the eigen-state of the system Hamiltonian involving all scattering processes and ii) when $\ket{n}$ is the eigen-state of LLL Hamiltonian where all scattering processes are confined to LLL. The transition matrix element $I_n$ represents the system transition rate due to an oscillating metric tensor analogous to the passing of a gravitation wave and hence usually termed as \textit{graviton} \cite{kun_PRB}. Fig. ~\ref{fig:grav} (a) and (b) shows the presence of such mode(s) for various system sizes. This mode occurs for both liquid limit and near-TT limit (See SM \cite{supp}). As seen from Fig. \ref{fig:grav}, the energy of graviton states $E_{grav}$, when all scattering processes are included, is slightly lower than the graviton states of a system when all interaction processes are confined to LLL. Furthermore, we follow Ref.~\onlinecite{chiralRezayi} to analyze the handedness of this graviton and find that it has the structure of $(q_x-iq_y)^2$, suggesting its chiral nature (See SM ~\cite{supp} for the detail of analysis.). Previously, it has also been proposed ~\cite{AmmarPRL, PapicMain} that this low-energy mode can be excited via geometric quench \cite{PapicBilayer}. A variant of such graviton mode called chiral graviton mode has been observed in experiments recently for fractional Hall states at filling fractional $\frac{1}{3}$ and $\frac{2}{5}$ fillings \cite{NatureGraviton}. Chiral graviton modes in FQH system has been previously excited by tuning the laser-frequency near resonance conditions between valence and conduction band in 2D quantum wells. Our work has demonstrated that chiral graviton modes can also be exited when the a short-time laser is tuned near Landau level gap which is easily accessible experimentally. Furthermore, the excitation of graviton mode can be observed along with magneto-plasmon mode when all scattering processes are considered as seen from the structure factor.  
\begin{figure}
\centering
  \includegraphics[width=1.0\linewidth]{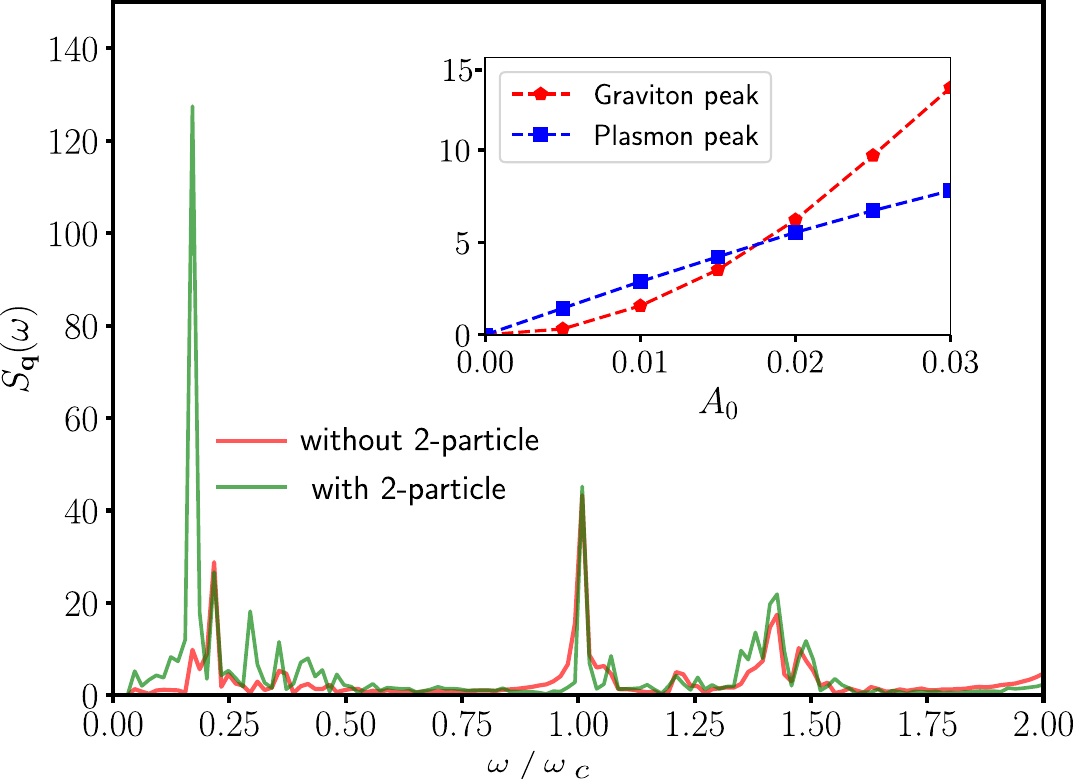}
  \caption{The structure factor $S_\mathbf{q}(\omega)$ (unnormalized) for $\mathbf{q}=(2\pi/L_x,0)$, $\omega=0.7\omega_c$ and $N=6$ electron system. All parameters are set according to the Fig.~\eqref{fig:sq}. Without the two-particle chiral excitations (Fig.~\ref{fig:excite} (c)) the intra-LL mode is suppressed. Inset gives the strength of the main peak of the low-energy (graviton) and plasmon mode against the intensity of the laser-pulse.  Quadratic dependence and linear dependence is evident for the low energy mode and plasmon mode respectively. The peak value of the plasmon mode is divided by 5 for better fit.}
  \label{fig:benidea}
\end{figure}
 The appearent drop in the dynamical structure factor when normalized in Fig.~\ref{fig:sq} between $N=6$ and $N=7$ is due to the presence of more quadrupole excitations present in $N=7$ photo-excited state $\ket{\Psi(t)}$. As seen from Fig.~\ref{fig:grav} (a) and (b), there are more graviton states for $N=7$ consistent with Ref. \onlinecite{chiralRezayi}. These states are also identified in Table.~\ref{tab:grav_updated}. Most importantly, the total low-energy quadrupole overlap in the photo-induced state does not drop with the system size as seen from Table.~\ref{tab:grav_updated}. We have shown that the quadrupole excitations occur up-to $N=8$. 

The two-particle process that excites the low-energy mode(s) through photo excitation, is given in Fig.~\ref{fig:excite} (c). This process describes inter-level two-electron transitions from one Landau level into another thus requiring two-photons. Moreover, this process has amplitude that is chiral in nature $\frac{(q_x-iq_y)^2}{2\sqrt{q_x^2+q_y^2}}e^{-\frac{q^2}{2}}c_{1,j_1}\dg c_{1,j_2}\dg c_{0,j_3}c_{0,j_4}+\mathrm{H.c.}$ (See SM \cite{supp}). As seen in Fig.~\ref{fig:benidea}, when such (chiral) two-particle excitations are omitted, there is a significant drop in the low-energy mode present in the structure factor. The dependence of the low-energy and the plasmon peak strength with the laser intensity in the limit $A_0\to 0$ is also shown in inset of Fig.~\ref{fig:benidea}. It is evident that in the low intensity regime, the magnetoplasmon and the low-energy mode follow the linear and quadratic dependence of the field strength respectively. This is in accordance with the fact that graviton couple with two-particle processes while magnetoplasmon depends upon single-photon process. Quantifying such quadratic dependence gives an alternate avenue to measure the graviton modes through non-linear absorption experiments. One can then use a circularly polarized probe light to experimentally detect the chirality of the excited graviton modes if needed. 
 
{\it Conclusions.---} Through exact-diagonalization, the laser-induced dynamics of FQH state at $\nu=1/3$ has been presented. We showed that laser-induced excitations of FQH states have rich physics. On the one hand, some excitations are coupled to electromagnetic field and can cause emission like the plasmon mode. On the other hand, we have shown that when scattering from LLL to LL1 is included in interactions, it is possible to excite neutral-modes of LLL through linearly polarized laser pulse field. These excitations are analogous to emergent \emph{gravitons} due to their quadrupole structure and is the excitation of intrinsic metric carried by Hall fluids. Moreover, we have shown that non-linear light absorption experiments are an alternate route to the detection of such modes being quadratically dependent upon the light's intensity. Future non-linear absorption experiments are needed to probe such excitations. This can also help with understanding how the gap structure for fractional Hall systems evolve with finite momentum. Such experiments will not only address the questions related to strongly-interacting electron systems, but can also shed light on the understanding of the analogues of fundamental particles in high energy physics realized in condensed matter systems.

{\it Acknowledgements.---}
This work was carried out under the auspices of the U.S. Department of Energy (DOE) National Nuclear Security Administration (NNSA) under Contract No. 89233218CNA000001. It was supported by Quantum Science Center, a U.S. DOE Office of Science Quantum Information Science Research Center, and in part by Center for Integrated nanotechnologies, a DOE BES user facility, in partnership with the LANL Institutional Computing Program for computational resources. A.K. thanks Avadh Saxena and Pouyan Ghaemi for their fruitful discussions. A.K. was supported LANL LDRD Program.

\bibliographystyle{apsrev4-1}

\bibliography{FQHE.bib}

\onecolumngrid 
\newpage

\end{document}


\title{Laser-driven Ultrafast Dynamics of a Fractional Quantum Hall System}

\author{Ammar Kirmani} 
\affiliation{Theoretical Division, Los Alamos National Laboratory, Los Alamos, New Mexico 87545, USA}
\email{akirmani@lanl.gov}

\author{Benedikt Fauseweh}
\affiliation{Department of Physics, TU Dortmund University, 44227 Dortmund, Germany}
\email{benedikt.fauseweh@tu-dortmund.de}

\author{Jian-Xin Zhu}
\affiliation{Theoretical Division, Los Alamos National Laboratory, Los Alamos, New Mexico 87545, USA}
\affiliation{Center for Integrated Nanotechnologies, Los Alamos National Laboratory, Los Alamos, New Mexico 87545, USA}
\email{jxzhu@lanl.gov}

\maketitle
\section{Hamiltonian}
\noindent We truncate our model to the first Landau level and define $q=q_x-iq_y$ 
\begin{eqnarray}\label{eq:cc}
\nonumber
C_{n_1,n_4}(\bq)&=\sqrt{\frac{n_4!}{n_1!}}\bigg(\frac{iq\ell_B}{\sqrt{2}}\bigg)^{n_1-n_4} L_{n_4}^{n_4-n_1}\bigg(\frac{(q_x^2+q_y^2)\ell_B^2}{2}\bigg), \quad n_1 \geq n_4\\C_{n_1,n_4}(\bq)&=\sqrt{\frac{n_1!}{n_4!}}\bigg(\frac{iq^*\ell_B}{\sqrt{2}}\bigg)^{n_1-n_4} L_{n_4}^{n_1-n_4}\bigg(\frac{(q_x^2+q_y^2)\ell_B^2}{2}\bigg), \quad n_1 \leq n_4
\end{eqnarray}

\begin{eqnarray}\label{eq:H2}
V_{j_1,j_2,j_3,j_4}^{n_1,n_2,n_3,n_4}=\delta'_{j_1+j_2,j_3+j_4}\frac{1}{2L_1 L_2}\sum_{\bq=\big(s\frac{2\pi}{L_1},t\frac{2\pi}{L_2}\big)}^{'}\delta'_{j_1-j_4,t} V(\bq)C_{n_1,n_4}(-q_x,-q_y)C_{n_2,n_3}(q_x,q_y)e^{-i2\pi s(j_1-j_3)/N_{\phi}}e^{-\frac{q^2}{2}\ell_B^2}
\end{eqnarray}
Where $\delta'$ is defined modulo $N_\phi$ and $\bq'$ represents that $\bq=0$ is eliminated. $V(\bq)=\frac{2\pi}{\eps q}$ for Coulumb and $V(\bq)=-q^2$ for Haldane pseudo-potential.
\subsection{Projected density operators and structure factor}
\noindent The single-particle wavefunction for an electron in Landau orbital (site) $j$ and level $n$ for torus boundary conditions is given as
\begin{eqnarray}\label{eq:sp_orb}
\phi_{n,j}(\br)=\frac{1}{\sqrt{N_n}}\sum_l e^{iy X_{j,l} }e^{-\frac{(x+X_{j,l})^2}{2}}H_n \bigg(\frac{x+X_{j,l}}{\ell_B} \bigg)
\end{eqnarray}
Where $X_{j,l}= \frac{2\pi \ell_B^2}{L_y}j+ l L_x$. We set $\ell_B==1$ and define $\kappa=\frac{2\pi}{L_y}$. The overlap integrals can be given
\begin{eqnarray}\label{eq:denq}
&&I_{j,k}^{n,n'}(\bq)=\frac{1}{\sqrt{N_n N_n'}} \int d\br e^{i\bq.\br}\phi_{n,j}^*(\br)\phi_{n',k}(\br)\\
&&I^{0,0}_{j,k}(\bq)=\sum_{\Delta} \delta_{\frac{q_y}{\kappa},(j-k)+\Delta N_\phi}e^{-\frac{\bq^2}{4}}e^{-i\frac{q_x}{2}\big(X_j+X_k+\Delta L_x\big)}\nonumber\\&&I^{1,0}_{j,k}(\bq)=\frac{1}{\sqrt{2}}\sum_{\Delta} \delta_{\frac{q_y}{\kappa},(j-k)+\Delta N_\phi}\bigg(q_y+iq_x \bigg)e^{-\frac{\bq^2}{4}}e^{-i\frac{q_x}{2}\big(X_j+X_k+\Delta L_x\big)}\nonumber\\&&I^{0,1}_{j,k}(\bq)=\frac{1}{\sqrt{2}}\sum_{\Delta} \delta_{\frac{q_y}{\kappa},(j-k)+\Delta N_\phi}\bigg(-q_y+iq_x \bigg)e^{-\frac{\bq^2}{4}}e^{-i\frac{q_x}{2}\big(X_j+X_k+\Delta L_x\big)}\nonumber\\&&I^{1,1}_{j,k}(\bq)=\frac{1}{2}\sum_{\Delta} \delta_{\frac{q_y}{\kappa},(j-k)+\Delta N_\phi}\bigg(2-\bq^2\bigg)e^{-\frac{\bq^2}{4}}e^{-i\frac{q_x}{2}\big(X_j+X_k+\Delta L_x\big)}\nonumber
\end{eqnarray}
Using the above equations, we can evaluate the expression for the projected densities as
\begin{eqnarray}\label{eq:denq}
&&\hat{\rho}^{0,0}_\bq=\sum_{\Delta,j,k} \delta_{\frac{q_y}{\kappa},(j-k)+\Delta N_\phi}e^{-\frac{\bq^2}{4}}e^{-i\frac{q_x}{2}\big(X_j+X_k+\Delta L_x\big)}c_{0,j}\dg c_{0,k}\nonumber\\&&\hat{\rho}^{1,0}_\bq=\frac{1}{\sqrt{2}}\sum_{\Delta,j,k} \delta_{\frac{q_y}{\kappa},(j-k)+\Delta N_\phi}\bigg(q_y+iq_x \bigg)e^{-\frac{\bq^2}{4}}e^{-i\frac{q_x}{2}\big(X_j+X_k+\Delta L_x\big)}c_{1,j}\dg c_{0,k}\nonumber\\&&\hat{\rho}^{0,1}_\bq=\frac{1}{\sqrt{2}}\sum_{\Delta,j,k} \delta_{\frac{q_y}{\kappa},(j-k)+\Delta N_\phi}\bigg(-q_y+iq_x \bigg)e^{-\frac{\bq^2}{4}}e^{-i\frac{q_x}{2}\big(X_j+X_k+\Delta L_x\big)}c_{0,j}\dg c_{1,k}\nonumber\\&&\hat{\rho}^{1,1}_\bq=\frac{1}{2}\sum_{\Delta,j,k} \delta_{\frac{q_y}{\kappa},(j-k)+\Delta N_\phi}\bigg(2-\bq^2\bigg)e^{-\frac{\bq^2}{4}}e^{-i\frac{q_x}{2}\big(X_j+X_k+\Delta L_x\big)}c_{1,j}\dg c_{1,k} 
\end{eqnarray}
Given the above relations in Eq. (\ref{eq:denq}), we define
\beg
\hat{\rho}_{\bq}=\hat{\rho}_{\bq}^{0,0}+\hat{\rho}_{\bq}^{1,0}+\hat{\rho}_{\bq}^{0,1}+\rho_{\bq}^{1,1}
\en
The structure factor is then given by,
\beg
{\mathcal S}_{\bq}=\hat{\rho}_{\bq}\hat{\rho}_{-\bq}
\en
It is not difficult to get $\rho_{-\bq}$ by inverting the momentum for example inverting momentum in $\rho^{0,0}_{\bq}$ gives $\hat{\rho}^{0,0}_{-\bq}=\sum_{\Delta,j,k} \delta_{-\frac{q_y}{\kappa},(j-k)+\Delta N_\phi}e^{-\frac{\bq^2}{4}}e^{+i\frac{q_x}{2}\big(X_j+X_k+\Delta L_x\big)}c_{0,j}\dg c_{0,k}$.
\section{Low energy quadrupole excitations and handedness}
\noindent We follow \cite{chiralRezayi} in this section. The expression of the low-energy (neutral) quadrople exciation is given as  
\begin{eqnarray}\label{eq:grav}
\hat{O}^{(2)}=\sum_{\{j_i\}}\sum_{q=(s \frac{2\pi}{L_x},t\frac{2\pi}{L_y})}^{'}  \delta_{j_1+j_2,j_3+j_4}^{'}\delta_{j_1-j_4,t}^{'}\bigg(q_x^2-q_y^2\bigg)V_{\bq}e^{-\frac{q^2}{2}}e^{-i2\pi s\frac{j_1-j_3}{N_\phi}} c_{0,j_1}\dg c_{0,j_2}\dg c_{0,j_3}c_{0,j_4}
\end{eqnarray}
Where prime over summation means that $q=0$ is excluded and prime over delta functions means that they are defined modulo $N_\phi$. The chiral variant of Eq. (\ref{eq:grav}) given in Ref. \onlinecite{chiralRezayi}.
\begin{eqnarray}\label{eq:cgrav}
\hat{O}^{(2)}_{\mp}=\sum_{\{j_i\}}\sum_{q=(s \frac{2\pi}{L_x},t\frac{2\pi}{L_y})}^{'}  \delta_{j_1+j_2,j_3+j_4}^{'}\delta_{j_1-j_4,t}^{'}\bigg(q_x\mp iq_y\bigg)^2\frac{2\pi}{q}e^{-\frac{q^2}{2}}e^{-i2\pi s\frac{j_1-j_3}{N_\phi}} c_{0,j_1}\dg c_{0,j_2}\dg c_{0,j_3}c_{0,j_4}
\end{eqnarray}

\begin{figure}[H]
  \includegraphics[width=.8\linewidth]{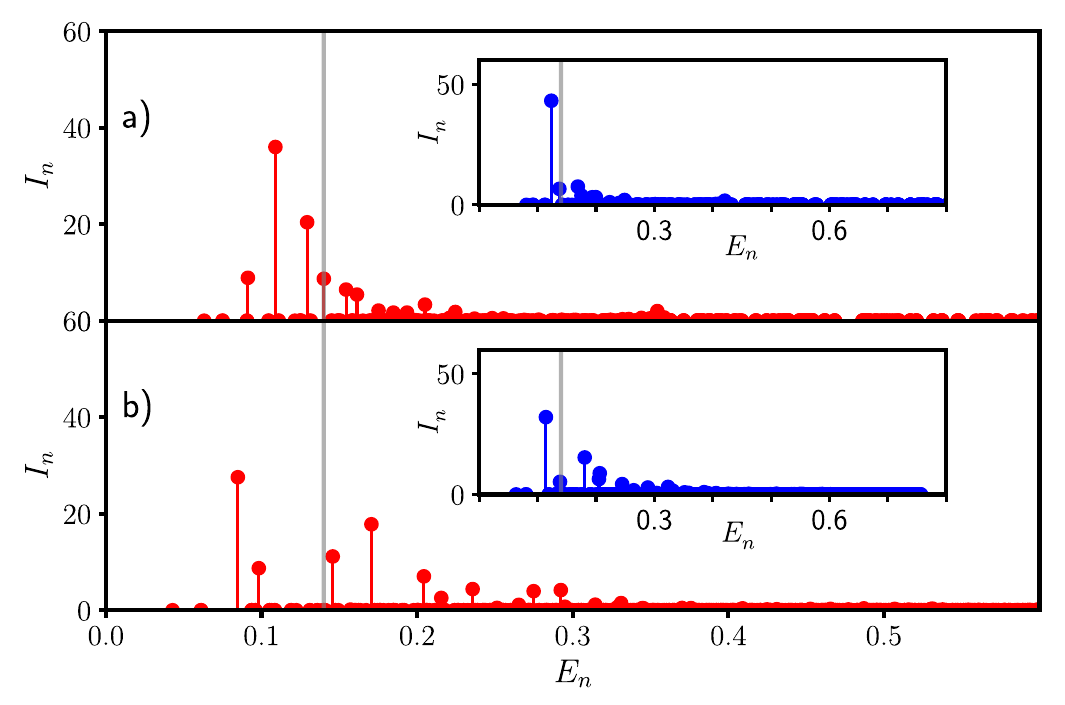}
  \caption{ Color online. (a) $I_n=|\bra{\Psi_0} \hat{O}^{(2)}\ket{n}|$ for near-TT limit $L_y=2\pi \ell_B$. Where $\ket{\Psi_0}$ is the FQH-1/3 state and $\ket{n}$ is the $n$-th excited state of the Hamiltonian involving all scattering processes. The inset of (a) gives the transition matrix element between FQH-1/3 state and the eigen-states of LLL. (b) Same as (a) but for quantum-liquid limit ($L_y=10.63\ell_B$) for square torus. The inset of (b) gives the matrix element for the liquid-limit in LLL. The system size is $N=6$ particles and the level-spacing is set to $0.5E_c$. Grey vertical line on plots is at energy $0.14 E_c$.}
  \label{fig:apn:grav}
\end{figure}

\begin{figure}[H]
\centering
  \includegraphics[width=0.8\linewidth]{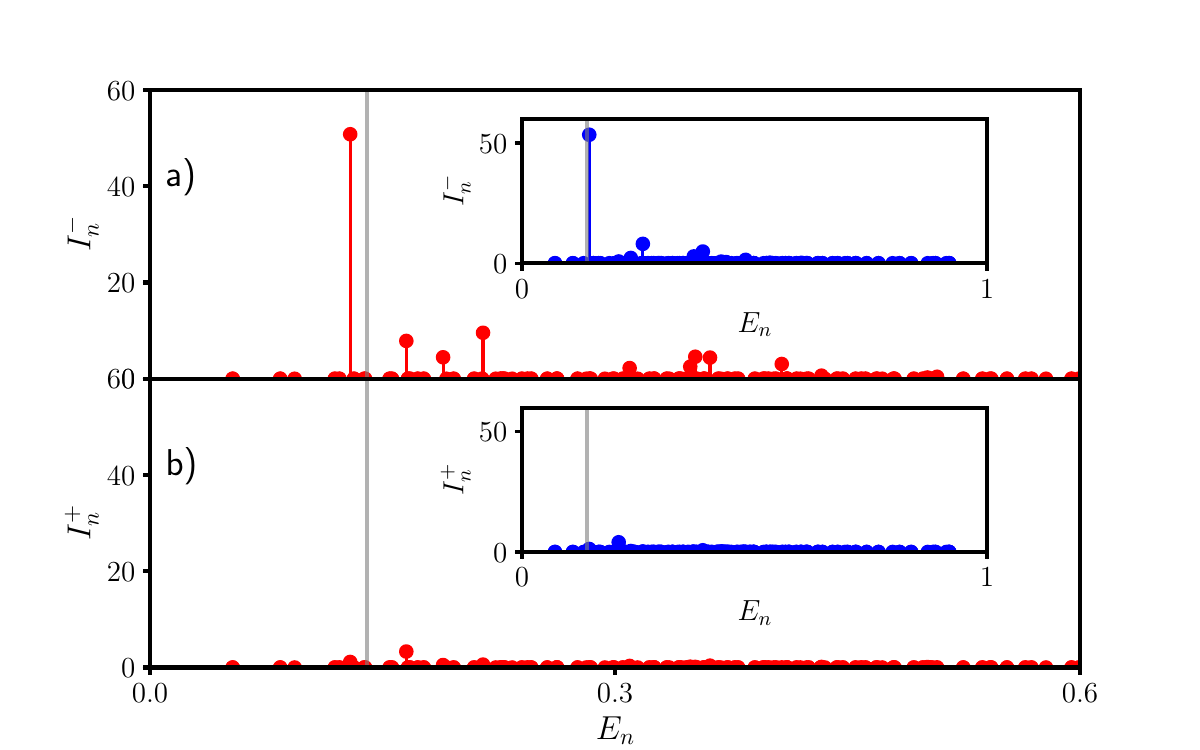}
  \caption{Color online. (a) $I_n^{-}=|\bra{\Psi_0} \hat{O}^{(2)}_{-}\ket{n}|$ for $N=5$ and $L_y=\sqrt{2\pi 15} \ell_B$. Where $\ket{\Psi_0}$ is the FQH-1/3 state and $\ket{n}$ is the n-th excited state of Hamiltonian with all inter- and inta-level interaction. The inset of (a) gives the transition matrix element between FQH-1/3 state and the eigen-states of LLL. (b) $I_n^{+}=|\bra{\Psi_0}\hat{O}^{(2)}_{+}\ket{n}|$ showing the suppression with all parameters are kept same as (a). The level-spacing is set to $0.5E_c$. Grey vertical line on plots is at energy $0.15 E_c$. $V_{\bq}=\frac{2\pi}{\eps q}$ is used. }
  \label{fig:apn:cgrav}
\end{figure}
From figure \ref{fig:apn:grav} and \ref{fig:apn:cgrav}, we can conclude that our state our graviton excitations are also chiral as agreeing with Ref. \onlinecite{chiralRezayi}.

\begin{figure}[H]
\centering
  \includegraphics[width=.6\linewidth]{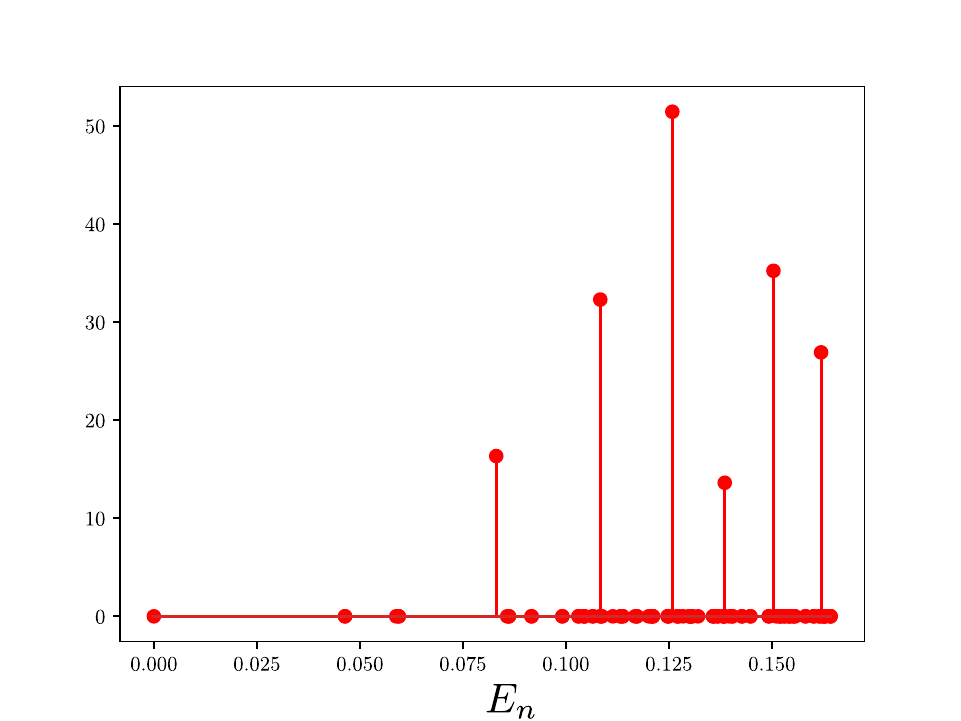}
  \caption{ Color online. $I_n=|\bra{\Psi_0} \hat{O}^{(2)}\ket{n}|$ for $N=8$ for $L_y=\sqrt{6\pi8} \ell_B$ only 100 lowest energy eigenstates were available. }
  \label{fig:apn:g8}
\end{figure}
\section{Excitation mechanism of quadrupole excitations}
\begin{figure}
\centering
  \includegraphics[width=0.75\linewidth]{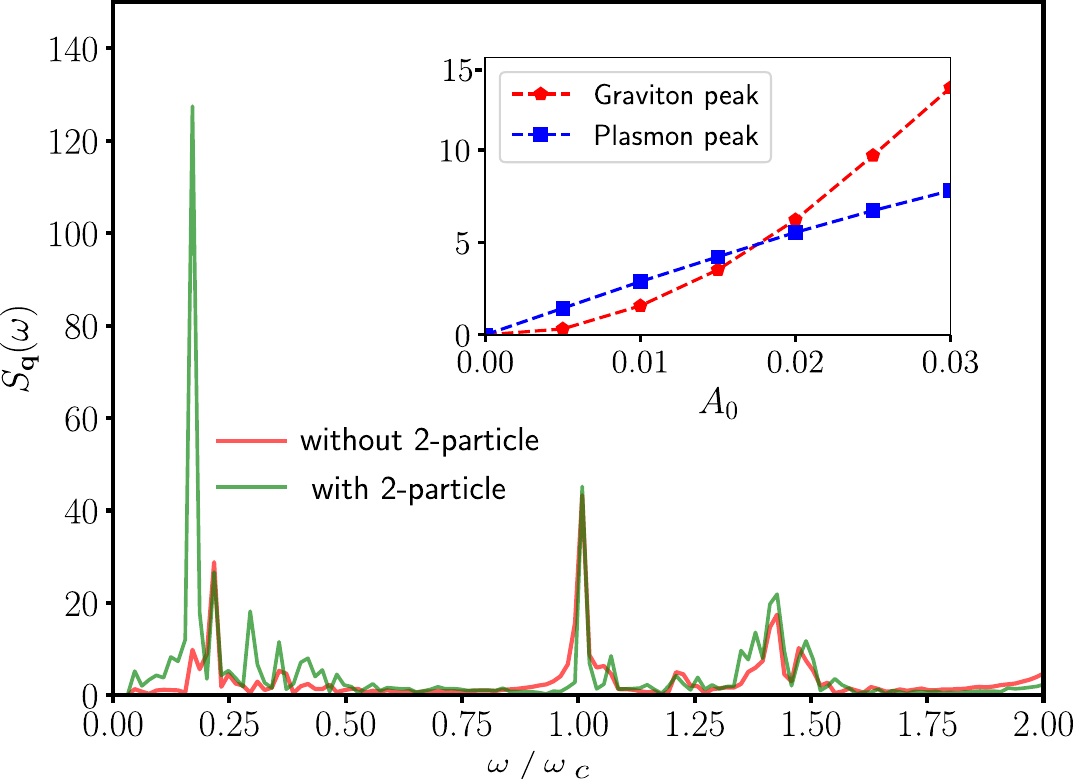}
  \caption{The structure factor $S_\mathbf{q}(\omega)$ (un-normalized) for $\mathbf{q}=(2\pi/L_x,0)$ and $N=6$ electron system. All parameters are set according to the Fig. (5) of the main text. Without the two-particle chiral excitations the intra-LL  mode is suppressed.}
  \label{fig:benidea}
\end{figure}

\begin{figure}
\centering
  \includegraphics[width=0.75\linewidth]{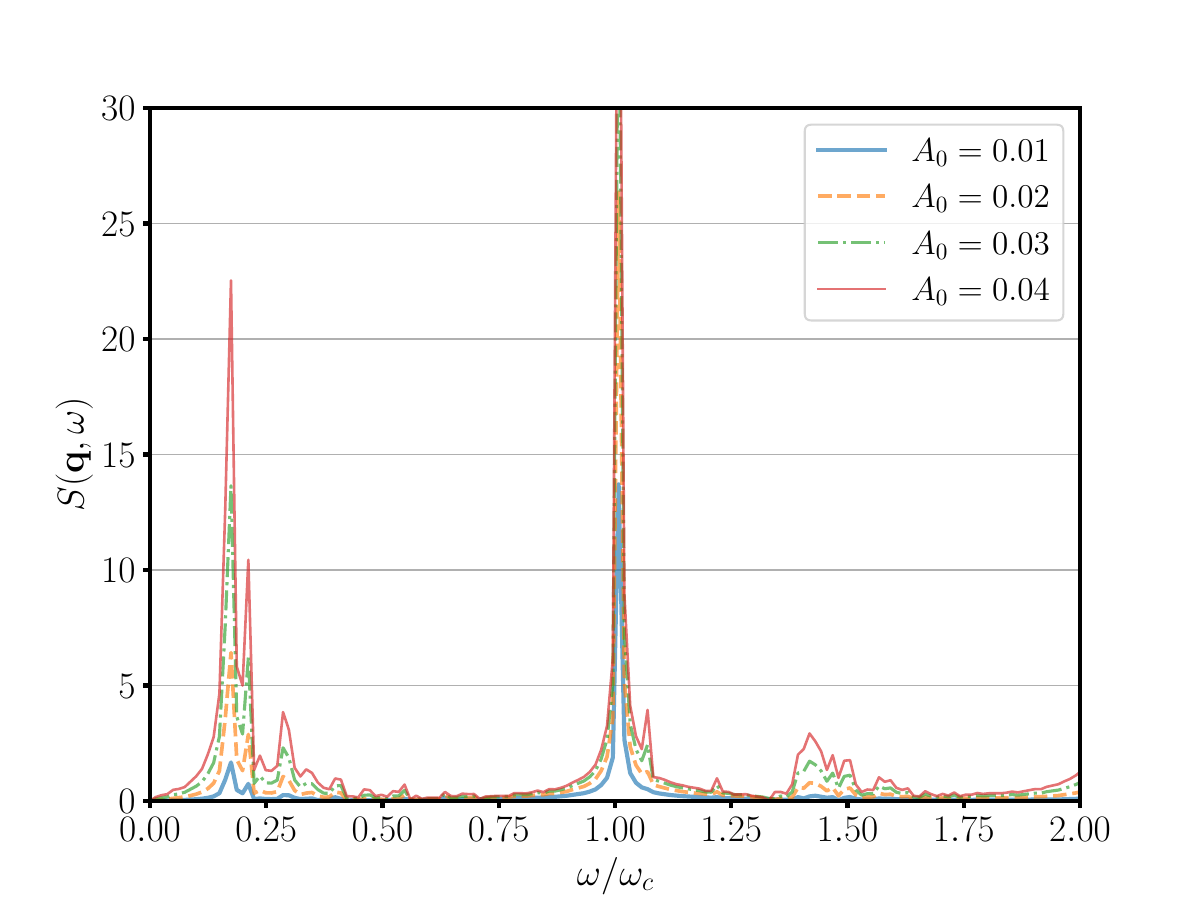}
  \caption{The structure factor $S_\mathbf{q}(\omega)$ $\mathbf{q}=(2\pi/L_x,0)$for $N=6$ electron system. The frequency of the laser-drive's is fixed at the half of Landau-level energy spacing $\omega_c/2$. }
  \label{fig:reply_strength}
\end{figure}

We have analyzed and found that the low-energy mode is mainly excited due to the following (in order)
\begin{enumerate}
    \item The incoming laser-drive excite single electron from $LLL$ to $LL1$. This is a single particle process couples with the light field.
    \item Once electrons are pumped in the $LL1$. There are chiral two-particle excitation processes between $LL1$ and $LLL$ that were previously ignored are proportional to \\$H_{\mathrm{chiral}} = \sum_{\{j_i\}}\sum_{q_x=k_1\frac{2\pi}{L_x},q_y=k_2\frac{2\pi}{L_y}}\frac{(q_x-iq_y)^2}{2\sqrt{q_x^2+q_y^2}}e^{-\frac{q^2}{2}}e^{\frac{-2\pi i k_1(j_1-j_3)}{N_\phi}}c_{1,j_1}\dg c_{1,j_2}\dg c_{0,j_3}c_{0,j_4}+\mathrm{H.c.}$
\end{enumerate}
We show that when such processes are considered, the excitation of intra-level mode is possible through two-electrons processes (See Fig. \ref{fig:benidea}). Moreover, our explanation predicts graviton response to be quadratically proportional to the field strength (See Fig. \ref{fig:reply_strength}) in agreement with the fact that graviton couples with two-photon process. Previous works [Phys. Rev. Lett. 119, 247403] and [Phys. Rev. Lett. 126, 076604] involving two Landau levels (or layers) keep the particle number fixed in each Landau level (or layer) and hence fail to include these interactions.

\begin{figure}
\centering
  \includegraphics[width=0.75\linewidth]{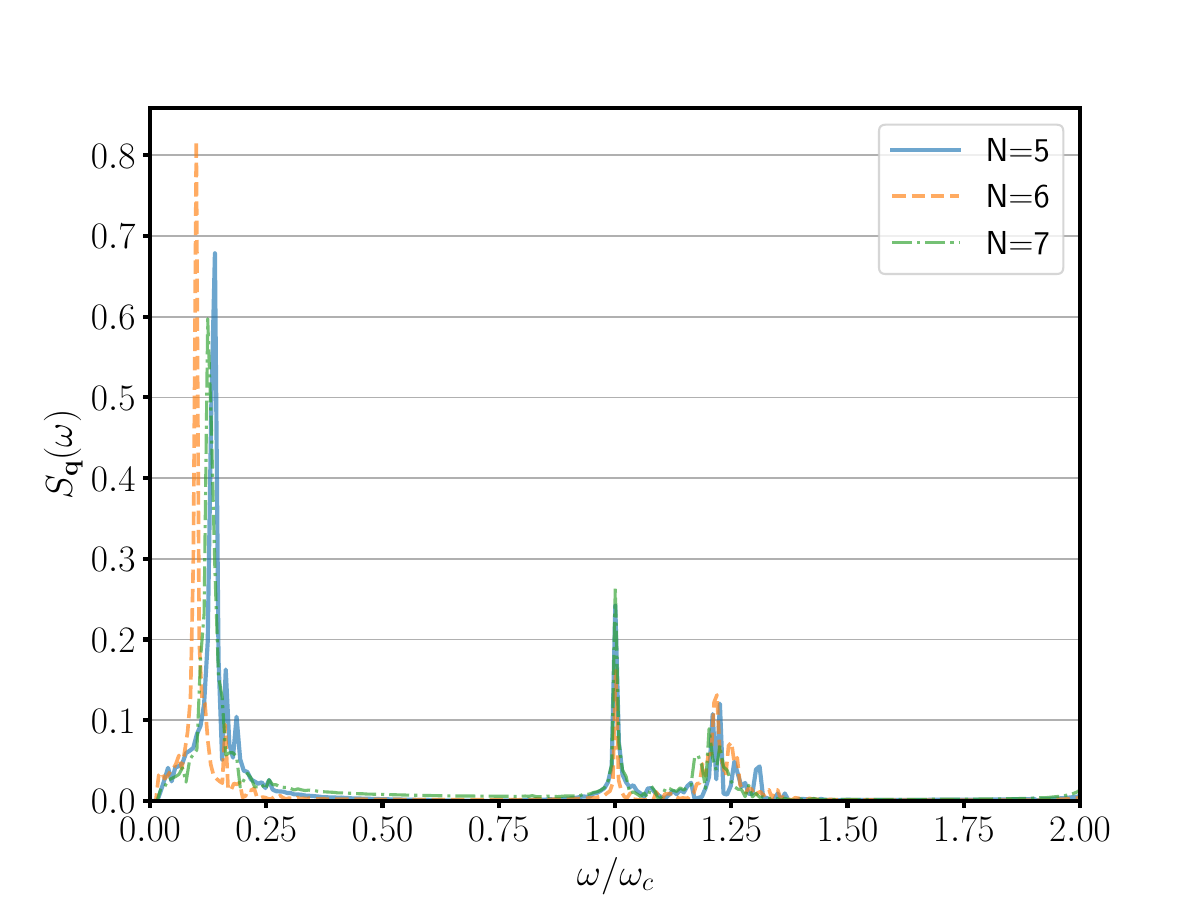}
  \caption{The structure factor $S_\mathbf{q}(\omega)$ for $\mathbf{q}=(2\pi/L_x,0)$for $\omega_c=E_c$ for $N=5,6,7$ electron system. $(eA_0/\sqrt{2}m^*\ell_B,\omega_0)=(0.4E_c,0.75\omega_c)$ with $t_d=10E_c$ and $L_y=\sqrt{2\pi N_\phi}\ell_B$.}
  \label{fig:reply_spacing}
\end{figure}
We present the results for larger-energy spacing results in Fig. \ref{fig:reply_spacing} showing the similar behavior (i.e. the excitation of intra-level low energy mode). We note that for $\omega_c\geq E_c$ the intra-level mode energy is almost independent of the level-spacing. The excitation of intra-level mode is possible for even larger $w_c$ as long as values of $\omega_0$ and laser-amplitude is adjusted. Generally in the regime $\omega\geq E_c$ the intra-level energy mode is dependent only on Coulumb energy scale $E_c$.
\section{Current}
The current is obtained by infinitesimally changing the gauge potential from $A=(0,Bx,0)$ to $A'$ in single-particle Hamiltonian and evaluating the difference $\hat{H}_{sp}(A')-\hat{H}_{sp}(A)$. Where $A'=A+\delta A_x \hat{x}$ for the x-component and $A'=A+\delta A_y \hat{y}$ for the y-component of the current respectively. The difference is then divided by $\delta A_x$ and $\delta A_y$. Arriving at
\begin{equation}
\label{eq:observables}
\hat{J}_{x(y)}=\frac{i(1)}{m\sqrt{2}}\sum_j \bigg(c_{1,j}\dg c_{0,j} \mp c_{0,j}\dg c_{1,j}\bigg)\;.
\end{equation}
\begin{figure}
  \includegraphics[width=1.0\linewidth]{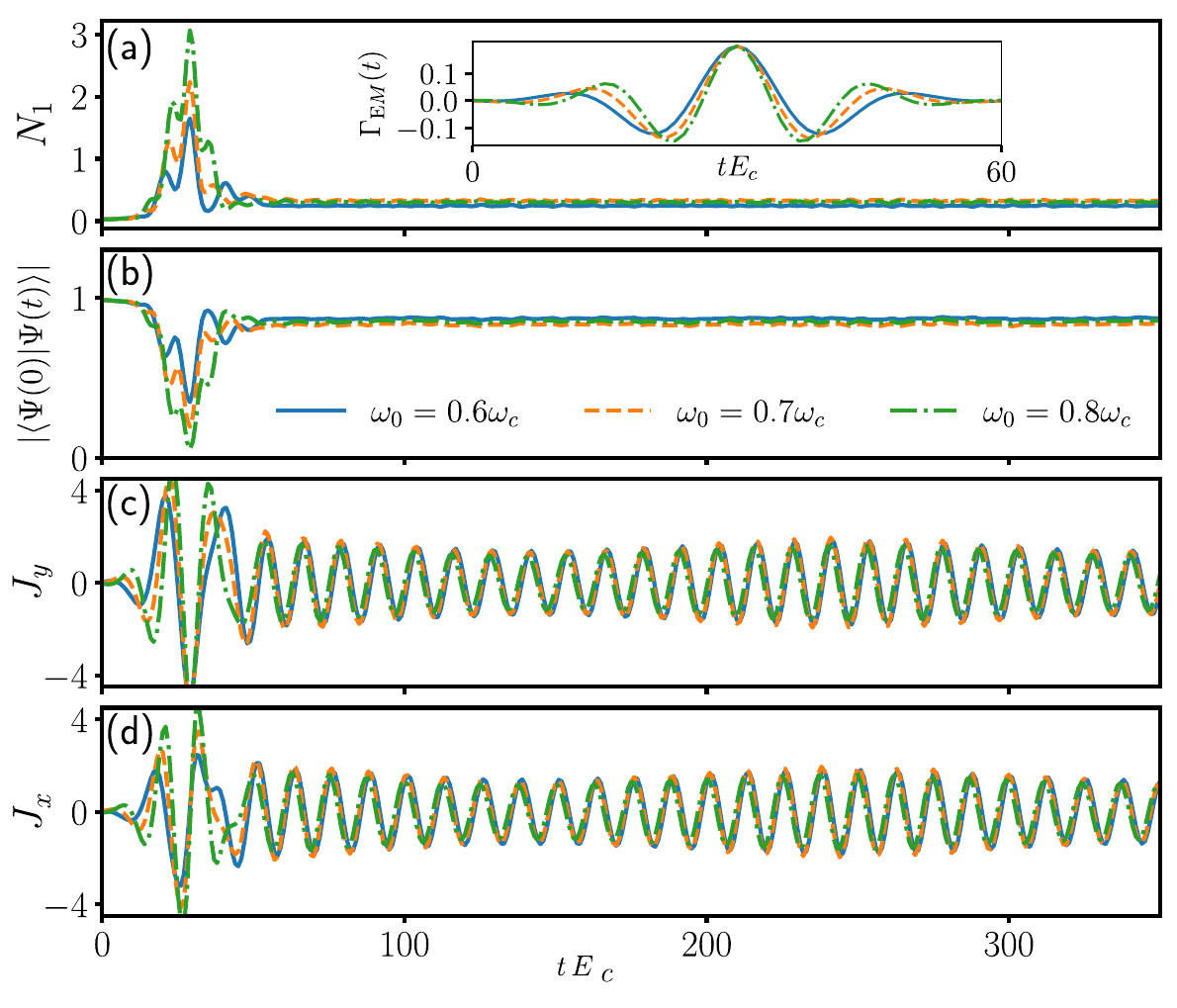}
  \caption{ Color online. Dynamical observables for various laser frequencies $\omega_0$ (a) Number of excited electrons for various in-coming light frequencies in-set of (a) gives the form of $\Gamma_{\rm EM}(t)$ (b) Fidelity $|\langle\Psi(0)\ket{\Psi(t)}|$. The quantities are obtained for $N=6$ on torus with $L_2=10.63\ell_B$. The intensity is set to $eA_0/\sqrt{2}m^*\ell_B=0.2E_c$ and $t_d=10/E_c$. The units of time is the inverse of $E_c$. The usual values of $E_c$ is around $14\; {\rm meV}$ \cite{inelastic1/3} which gives $47$ femto-seconds as the unit of time.}
  \label{fig:dynamics}
\end{figure}\\
We consider level-spacing $\omega_c<E_c$, i.e. the level spacing is less than Coulomb energy scale. The ratio of these energies $\kappa\equiv \frac{e^2}{\hbar \omega_c \epsilon\ell_B}$ is magnetic field dependent ranging from $2.6/\sqrt{B[T]}$ for Ga-As to $22.5/\sqrt{B[T]}$ for Al-As justifying the our approximation for fractional $\nu=1/3$ Hall system usually observed at $B \sim 10$ Tesla \cite{mixing2013}. In Fig. \ref{fig:dynamics}, we present observables like fidelity and the mean-excitation number. The current $J_{x(y)}$ in the Fig. \ref{fig:dynamics} (c-d) remains oscillating with multiple periods after the pulse is finished.

\begin{figure}
\includegraphics[width=1.0\linewidth]{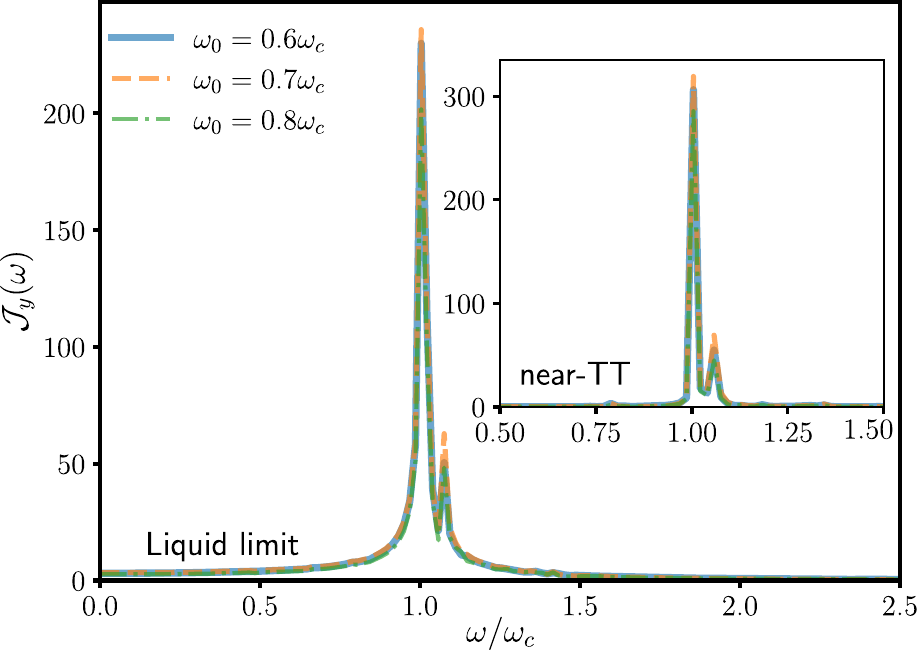}
\caption{ Spectrum of $d[\hat{J}_y(t)]/dt$ following the pulse with parameters $eA_0/\sqrt{2}m^*\ell_B=0.2E_c$ and $t_d=10/Ec$. The in-coming laser frequency is swept from $\omega_0=0.6-0.8\omega_c$. The behaviour of the current shows the excitation of a mode in addition to magnetoplasmon at frequency $w_c$ that is blue-shifted from $w_c$. The system size is set to $N=6$ and $L_y=\sqrt{2\pi 18}\ell_B$. Inset gives the spectrum for near-TT limit ($L_y=2\pi \ell_B$). }
 \label{fig:spectrum}
\end{figure}

\begin{figure}
\centering
 \includegraphics[width=\linewidth]{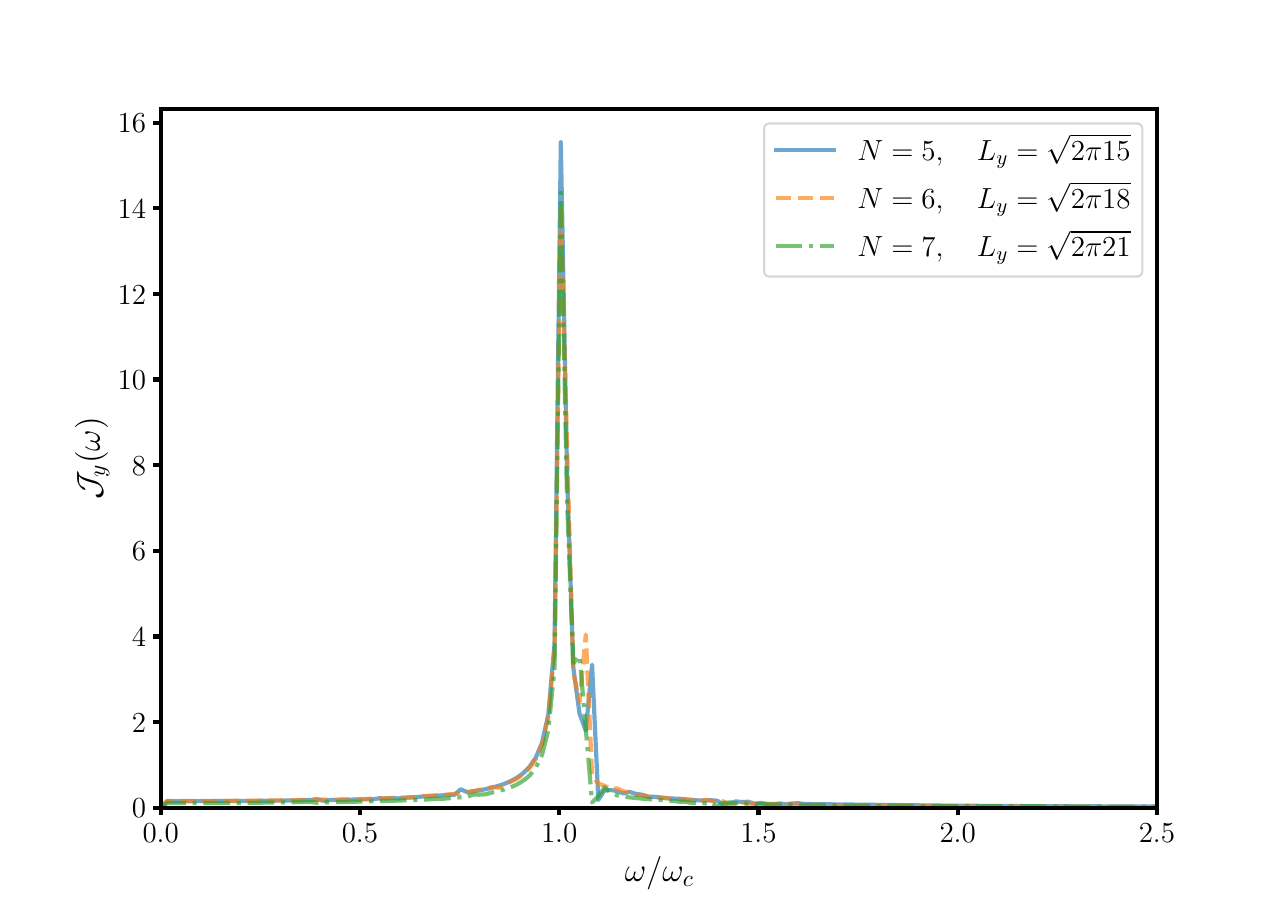}
\caption{Different runs for square torus $L_y=L_x=\sqrt{2\pi N_{\phi}}\ell_B$. The laser parameters are set to $(eA_0/\sqrt{2}m^*\ell_B,\omega_0)=(0.2,0.7\omega_c)$ and $\omega_c=0.5E_c$.}
\label{fig:appn:current}
\end{figure}

 In Fig. \ref{fig:dynamics}, we present observables like fidelity and the mean-excitation number. Fig. \ref{fig:dynamics} (a-b) gives the excitations to LL1 and the fidelity. The current $\hat{J}_{x,y}$ in the Fig. \ref{fig:dynamics} (c-d) remains oscillating with multiple periods after the pulse is finished. Fig. \ref{fig:spectrum} gives the spectrum of the derivative current ${\mathcal J}_{y}(\omega)={\mathcal F}_t \big[\frac{d}{dt}\hat{J}_{y}(t) \big](\omega)$ for $L_y=\sqrt{2\pi 18}\ell_B$ (liquid limit) and $L_y=2\pi\ell_B$ (near-TT limit). There are two modes as seen from the spectrum for low-intensity short time pulses one at $\omega_c$ and another blue-shifted from $\omega_c$. The presence of these modes are robust against slight change in the incoming frequencies although their strength can change. The peak at the level-spacing (cyclotron-energy) $\omega_c$ is the plasmon equivalent of FQH state identified as $q\to0$ limit of \emph{magnetoplasmon} mode. This mode is due to single particle excitations from LLL to LL1 \cite{inelastic1/3}. 
In the experiments, another peak that is blue shifted from magnetoplasmon has also been observed \cite{inelastic1/3}. This mode was sometimes attributed to $q=0$ spin-flip inter-Landau-level excitation involving enhanced exchanged in spin-polarized state. We find blue-shifted peaks in our calculations are due to excitations of additional electron to $LL1$ although complete understanding is missing due to the presence of finite-size effects and the truncation to LL1. Although, there is no net current due to the conservation of particle number, there is still a local time-dependent distribution of current e.g. along x-axis representing density-oscillations. In this section we will present more runs e.g. Fig. \ref{fig:appn:current} shows runs sizes $N= 5, 6, 7$ for square torus.
\section{Further low energy physics}
\begin{figure}
  \includegraphics[width=\linewidth]{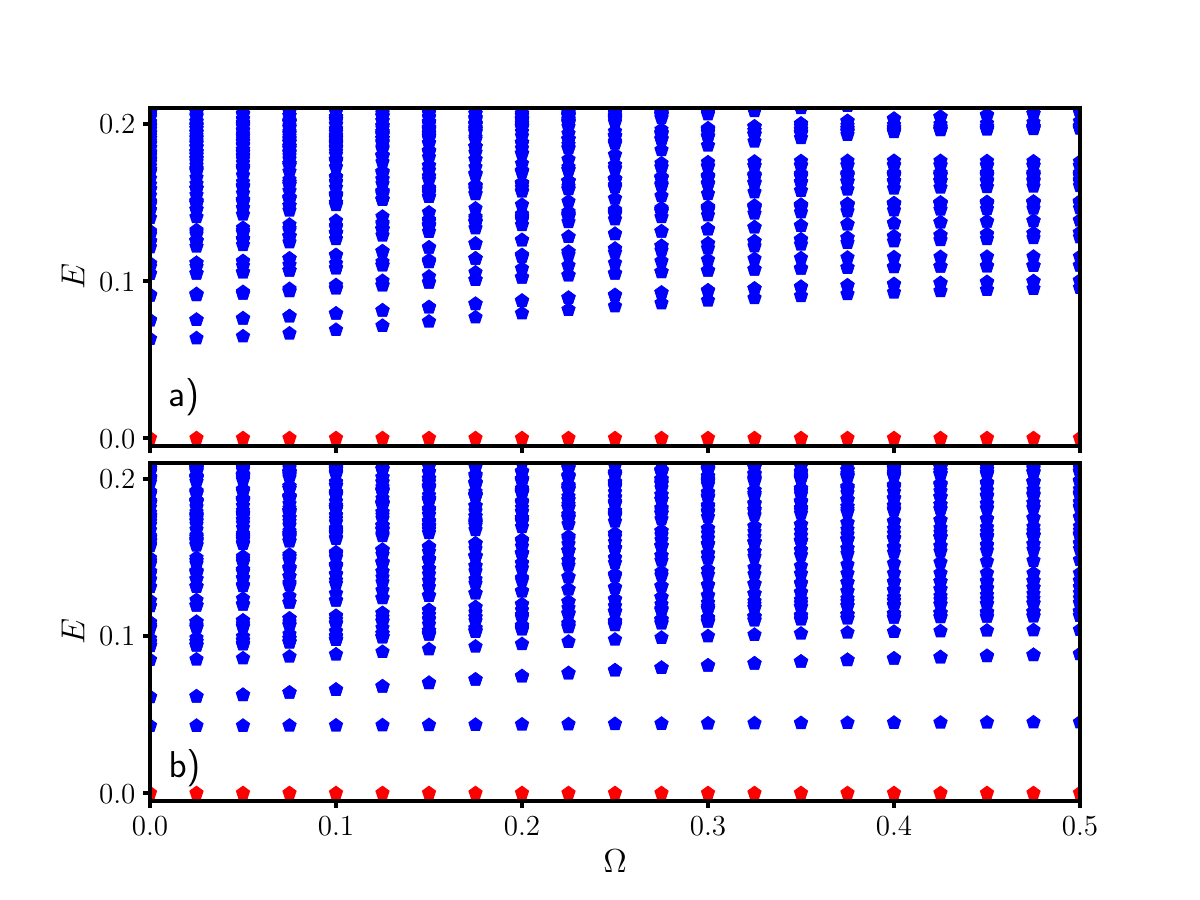}
  \caption{Color online. Low energy spectrum  of Eq. (\ref{eq:H2}) at $\nu=1/3$ for $\delta=0.5E_c$ at various values of $\Omega$.  (a) The circumference of the torus is set $b=2\pi \ell_B$ for near thin torus limit. (b) The circumference of the torus is set $b=\sqrt{2\pi 18} \ell_B$ for the liquid limit. The particle number is set to 6. The ground-state energy is shifted to 0 (red pentagons).}
  \label{fig:Low_energy}
\end{figure}
The equilibrium physics of Hamiltonian in Eq. (\ref{eq:H2}) can be explained by replacing $A_{EM}(t)$ by a static-field $\Omega$ that controls the amount of hopping between LLL and LL1. Fig. \ref{fig:Low_energy} shows the low-energy spectrum of Hamiltonian of Eq. (\ref{eq:H2}) at $1/3$ filling against various values of $\Omega$. The gap of ground state (G.S.) to the first excited state is robust for both near-TT limit (Fig. \ref{fig:Low_energy}-a) and the liquid-limit \ref{fig:Low_energy}-b). As $\Omega$ is increased, the excited state remains gapped from the ground state, although it is unclear how the G.S. for non-zero $\Omega$ is related to FQH $1/3$ state.\\
\begin{figure}
\centering
\includegraphics[width=0.8 \linewidth]{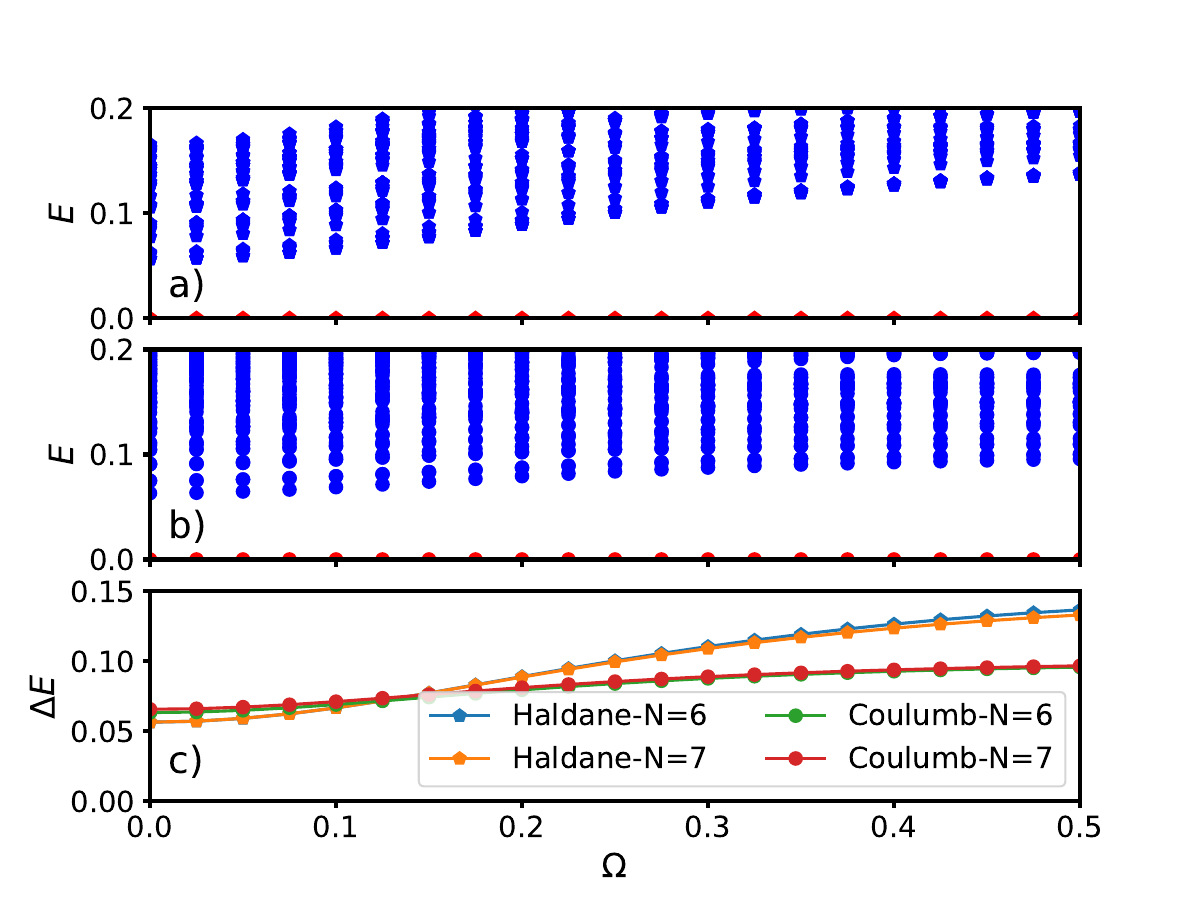}
\caption{ a) The low-energy spectrum of Haldane pseudo-potential Hamiltonian on torus for N=6 electrons at $1/3$ filling. b) Same as (a) but for Coulumb's potential. c) The energy gap to the excited state for different interactions and for various system sizes. The circumference of the torus is set to $L_y=2\pi \ell_B$.}
  \label{fig:apn:gap}
\end{figure}
To understand the low energy physics of our two-level model and its dynamics, we first establish the effect of inter-level two-body scattering for finite level spacing in the absence of the pulse field. For $V_1$ ($V_{\bq}\propto-q^2$) Haldane-pseudopotential the presence of inter-level scattering process do not affect the fractional $1/3$ Hall ground state. We evaluate the overlap of the ground state of our model with inter-level scattering processes with fractional Hall $1/3$ state. We have found that fractional Hall $1/3$ state for Haldane-pseudopotential is the ground state of our model even for small $\omega_c$.  Thus we can use our initial fractional $1/3$ Hall state to be the ground state of our system. For Coulumb potential on torus, the situation is similar but only at finite level-spacing $\omega_c$, the overlap with the fractional Hall 1/3 state recovers as shown in figure \ref{fig:apn:ovl} of Appendix. 
\begin{figure}
\centering
  \includegraphics[width=0.8\linewidth]{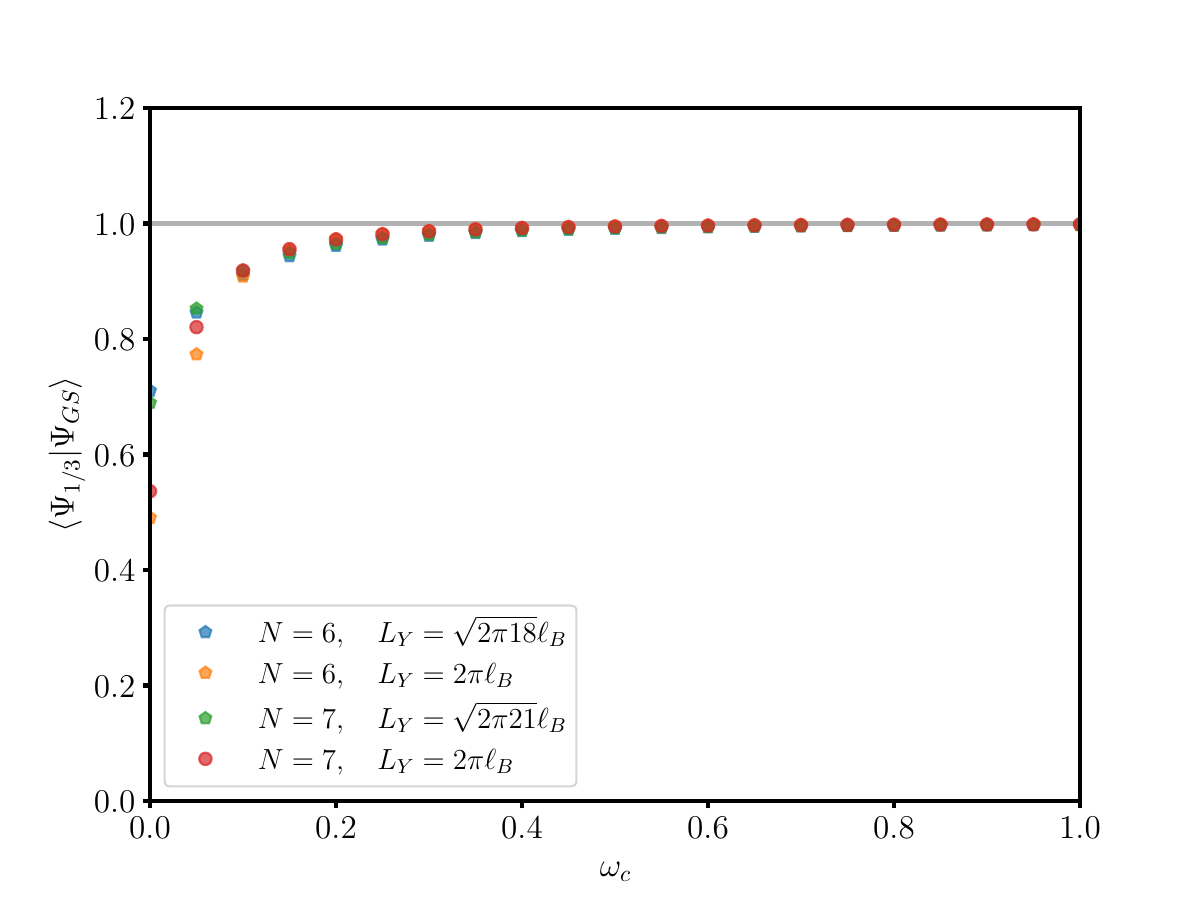}
  \caption{The overlap of fractional $1/3$ state in LLL with the ground state of Hamiltonian for Coulumb interactions on torus. The Hamiltonian has all matrix elements including Landau-level non-conserving interactions. $\omega_c$ is in the units of $e^2/\eps \ell_B$.}
  \label{fig:apn:ovl}
\end{figure}
We also plot the energy gap between the ground state and the first excited state on cylinder and torus for $V_1$ interactions in Fig. \ref{fig:apn:gap}. We see that for different system sizes, the system is gapped for all values of $\Omega_0$. The same hold true for Coulomb interactions on torus.
\bibliography{FQHE.bib}